# DATA CONFIDENTIALITY IN P2P COMMUNICATION AND SMART CONTRACTS OF BLOCKCHAIN IN INDUSTRY 4.0

Jan Stodt and Christoph Reich

Institute for Data Science, Cloud Computing, and IT Security at the University of Applied Sciences Furtwangen, Furtwangen, Baden-Württemberg, Germany

*ABSTRACT*

*Increased collaborative production and dynamic selection of production partners within industry 4.0 manufacturing leads to ever-increasing automatic data exchange between companies. Automatic and unsupervised data exchange creates new attack vectors, which could be used by a malicious insider to leak secrets via an otherwise considered secure channel without anyone noticing. In this paper we reflect upon approaches to prevent the exposure of secret data via blockchain technology, while also providing auditable proof of data exchange. We show that previous blockchain based privacy protection approaches offer protection, but give the control of the data to (potentially not trustworthy) third parties, which also can be considered a privacy violation. The approach taken in this paper is not utilize centralized data storage for data. It realizes data confidentiality of P2P communication and data processing in smart contracts of blockchains.*

*KEYWORDS*

*blockchain, privacy protection, P2P communication, smart contracts, industry 4.0*

## 1. INTRODUCTION

With the utilization of new technologies, business model and increased collaborative production leads to ever-increasing automatic data exchange between companies and service providers effective privacy protection is rendered more complicated. Previously in-house hosted services are outsourced to service providers with the goal of cost reduction and increased value creation. In the field of the manufacturing industry, companies may cooperate on producing a product together in a more dynamic way than ever before, resulting in the need for automated and privacy preserving P2P information exchange, as seen in Figure 1.

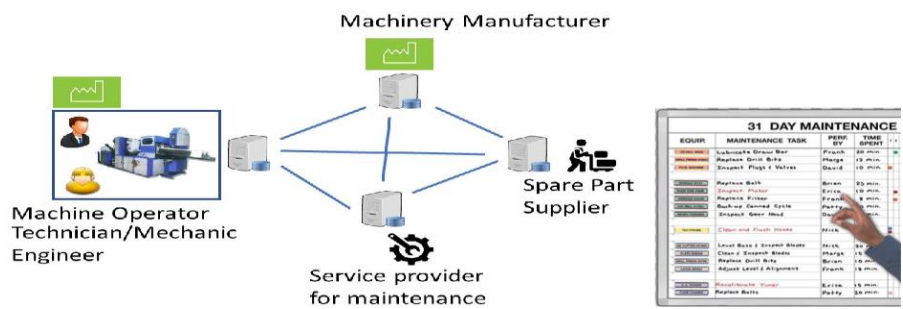

Figure 1. Peer-to-peer Communication in Industry 4.0; Maintenance Use Case

This paradigm shift is manifested by the concept of industry 4.0 [1]. Formerly data was processed on-premise. Today data is sent to external services and other partners for business automation. Examples ranging for process data exchange for production planning, quality data exchange for quality assurance, to exchange of maintenance logs for machine maintenance; all with the goal of increased service value while also reducing the cost for the service. Intensive communication between different industry partners of potentially private, sensitive, or confidential data results in loss of control over the data and may leading to serious privacy violations or data breaches. A recent example is the exposure of confidential data via a robotics vendor, as discussed by UpGuard [2], through an automated and uncontrolled data exchange. The automated data exchange led to exposure of confidential production information, drivers' licences of employees among other highly confidential data. This incident underlines the requirement for confidentiality and privacy protection in automated data exchange.

The companies are in a dilemma. On the one hand they want to automate the processes between their supplier and service provider, but on the other hand they do not want to reveal their crown jewels of data.

We provide an overview of the three most important state-of-the-art approaches of blockchain based confidentiality and privacy protection methods. We assess their approach of providing protection and compare them to our approach in the aspect of applicability for an industry 4.0 environment.

The main objectives of this paper are:

1. Evidence collection through blockchain, but remaining confidentiality of data: Blockchain is used to collect evidence of the automated business process for future audits. The developed Data Communication Module for Blockchain (DCMB) sends data by P2P communication directly to the participants and collects data exchange evidence.

2. Data confidentiality by using smart contracts: The developed Data Communication Module for Blockchain (DCMB) enables working with smart contracts using data signatures to keep the data confidential.

3. Data Protection Against Malicious Intrusion: The trusted execution environment (TEE) (see Sec. 3.3) protects against infrastructure attacks, with the goal of gaining access to private data.

## 2. RELATED WORK

Methods for privacy protection depend on the environment in which the data is located, the desired protection level, the use case of data processing, the potential privacy violates among other characteristics. *To the best of my knowledge, there are no papers discussing the application of blockchain technology for confidential and privacy protection in industrial partner collaboration.* Current Blockchain data confidentiality protection approaches are applied in medical data, voting and personal data storage. These approaches can be grouped into three categories: a) privacy preserving data mining b) data access control c) confidential smart contracts.

*Privacy Preserving Data Mining:* Privacy preserving data mining or privacy protecting computation, in combination with blockchain was delineated by Frey et al. and Zyskind et al. [3] among other similar concepts. All concepts share an analogous approach: the utilization of Multiparty Computation (MPC). With MPC it is possible to execute computations against data without direct data access. Not even the entities running the algorithm are able to extract data. To reduce the required storage volume for the data be analysed, the data is stored off-chain, meaning in an external storage system. Benhamouda et al. proposed a concept for implementing MPC in Hyperledger Fabric to support private data [4].

Privacy preserving data mining is a promising approach to confidentiality protection, as no direct access to data is possible. However, reflecting on the taxonomy of privacy by Solove [5], privacy preserving data mining protects against the privacy violation secondary use, but does not protect against exclusion and intrusion, thus the loss of data access if the external storage system is offline and therefore interruption of information flow. Loss of data access and the resulting interruption of data flow is a major problem for industrial partner collaborations, as it can lead to high financial losses, e.g. due to production line stops.

*Access Control:* Another approach to protect privacy is to specify, which entities may have access to data via blockchain based access control. Blockchain is hereby used to store the access rights and the location/address of the data in a secure, tamper-proofed and audit-able manner. Similar to privacy preserving data mining, data is stored off-chain in a centralized cloud storage system to reduce the storage volume requirements of the blockchain. The proposed concepts for this approach are varying in the granularity of the access policies. A notable concept in the area of healthcare is the paper of Yue et al. [6], which are proposing a concept of privacy-aware access policies, which are capable of fine-grained access control. In contrast to traditional access control models, which are only focused on who is performing an action on a data object, privacy-aware access policies are able to define rules regarding with which purpose data is accessed. Maesa et al. [7] propose publishing policies to the blockchain that are expressing the right to access a resource. The policies and the rights exchanges are publicly visible on the blockchain, thus any user knows at any given time the policy paired with a resource and the subjects who currently have the rights to access the resource.

*Smart Contract Privacy:* If data has to be processed confidentially and kept secret from the participants running the blockchain network, two approaches can be utilized. The work by Cheng et. al [8] developed confidentiality-preserving smart contracts, which are executed within a separate Trusted Execution Environments (TEEs). The blockchain is used to store an encrypted contract state. Hawk [9] is a smart contract system that provides confidentiality by executing contracts off-chain and posting only zero-knowledge proofs on-chain.

*Drawbacks of current approaches:* Privacy preserving data mining does not meet the specified requirement of P2P communication between industrial partners, as seen in Fig. 1. The concept of deriving information from data without direct access on raw data will be considered in our approach, albeit in a modified form. Access control does not allow the desired flexibility in the area of communication through P2P communication. Smart contract privacy is the most promising of the three approaches presented. However, the main requirement of the outlined industry 4.0 environment, data exchange between partners, is not provided by this approach. Smart contract privacy only led to the decision of "data is confidential" and "data is not confidential" in a non-observable way, access to cleared data is not provided by this approach. The non-observable smart contract execution approach will be considered in our approach, albeit with the addition of P2P data exchange between industrial partners.

## 3. DATA CONFIDENTIALITY IN BLOCKCHAINS FOR P2P COMMUNICATION AND SMART CONTRACTS

### 3.1. Confidentiality Audit Trail with Blockchain

The typical use of a blockchain is to collect data for giving evidence of a company-to-company communication (see Fig. 1), that has been sent correctly and in time. Often the data send between companies is confidential (e.g. number of parts being produced to calculate the leasing rate of a machine). In the case that all participants of the blockchain must be able to audit the data transfer between companies, but must not have access to the data, an auditable transfer log (audit trail) must be created. Fig. 2 shows, that data is transferred directly between machine owner and machine manufacturer.

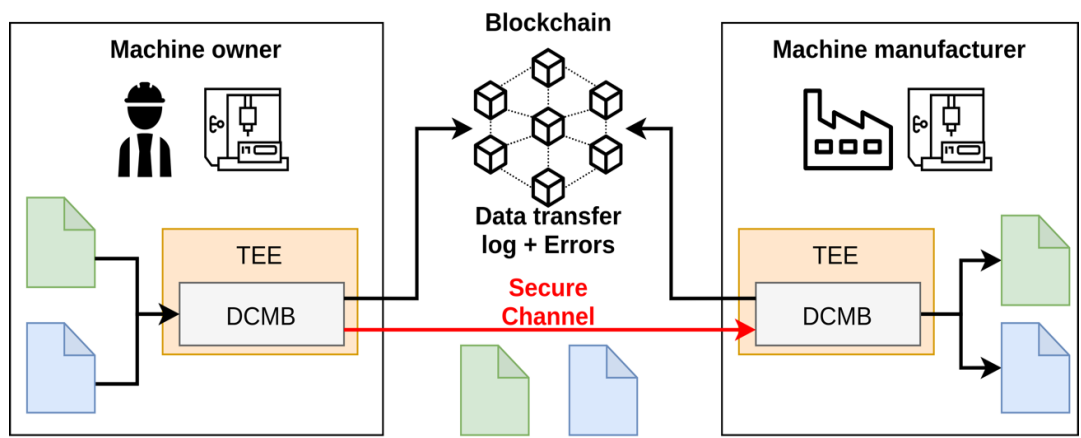

Figure 2. Audit Trail - Logging of Each Data Transfer

Direct streaming of data instead of storing the data into the blockchain also reduces the time between data sending, data clearing, and data receiving. In the case of unavailability, the data and the control over it can only be lost for a small amount of data. To protect data against tamper and unauthorized access at the location of sending and receiving, and during transport, a new intermediate module, *Data Communication Module for Blockchain (DCMB)* has been designed (see Fig. 2), that has the following functions:

- It collects data and sending it to the receiver via a secure channel,
- it provides a receiver with the streamed and cleared data,
- it sends evidence (message signatures) to the blockchain, to create an audit trail, and
- it signs collected messages to one blob and signs the total blob to reduce the number of evidence messages to the blockchain.

The module allows choosing individually at which location the privacy protection should take place, making this approach applicable for a wide variety of use cases, ranging for environments with limited resources such as IoT applications to applications with high resource requirements such as machine learning. To prevent extraction of the data, which has not yet been cleared and prevent tamper of the module, the module is deployed and executed in a trusted execution environment (TEE) (see Sec. 3.3 for more details). In essence, a TEE is an isolated environment that guarantees code and data loaded inside the environment to be protected against attacks in respect to confidentiality and integrity. Therefore, it can be assumed that either the module or the data can be tampered with. The use of the TEE aims to prevent access to uncleared data by attackers or the data recipients.

Algorithm 1 shows an example of the sender's communication. Within a given period of time, data is collected and the hash of the data is calculated with the hash function and a salt value. The hash and salt value of this algorithm, and all following algorithms, are generated by using the Argon2 algorithm [10]. The calculated hash and the salt for its creation is stored in the blockchain to provide an audit trail.

---
**Algorithm 1** Communication Node A (Sender)

1: **for each** time_interval **do**
1:    collect data;
1:    H = hash(data + salt);
1:    send: H + salt to Blockchain for audit trail
2: **end for**

---

Algorithm 2 shows an example for the communication of the receiver, input values are the hash and salt. $data1_{receiver}$ and $data2_{receiver}$ are examples for arbitrary data know to the receiver. Depending on which data, in combination with the received salt, is equal with received hash, different actions are performed.

---
**Algorithm 2** Communication Node B (Receiver)

1: input: H + salt
2: **if** H == hash($data1_{receiver}$ + salt) **then**
3:    action A (e.g. initiating ordering service);
4: **else if** H == hash($data2_{receiver}$ + salt) **then**
5:    action B;
6: **else if** ... **then**
7:    ...
8: **end if**

---

### 3.2. Confidentiality Smart Contracts

In the case that data must be recorded into the blockchain for processing in a smart contract, the DCMB maps the data into a qualitative description of an interval (e.g. "maintenance required in one {day, week, month}"). The mapped interval is then recorded into the blockchain. It must be noted that the mapping from quantitative values to a qualitative interval description reduces the possible functional scope of the smart contract: inequality, such as value comparisons, are no longer possible. The smart contract can therefore only check for the match of arbitrary conditions. Should it be necessary for the processing on the receiver's side to have the original data, it can also be sent directly to the receiver via P2P communication. An additional function of the DCMB is hashing of values before they are sent to the blockchain to keep the information confidential.

Algorithm 3 shows an example of a smart contract. If the transaction value (input of the smart contract) corresponds to a hash value stored in the blockchain, a certain action (e.g. send message) is executed by the smart contract.

**Algorithm 3** Smart Contract
1: **if** transaction_value == stored_hashed_value **then**
2:     action (e.g. send message);
3: **end if**

### 3.3. Data Communication Module for Blockchain Certification

To execute code in a secure, trusted and non-observable manner, a Trusted Execution Environment (TEE) may be used, which is based on a Secure Execution Environment (SEE). A SEE provides authenticity, integrity and confidentiality. In additional a TEE also provides remote attestation to proof its trustworthiness and must be resistant against attacks.

To provide the user of this approach with insights of the DCMB, the source code of the module should be published to the data sender, the data receiver and a group of validation entities for validation via a private repository shared by the mentioned entities. Should the source code contain private information, such in Algorithm 4, the source code is only published to the group of validation entities. This publication of the source code also enables governance and regulation of the privacy of the DCMB by a group of validation and certification entities for increased trust in this module. To ensure that source code and the module is mapped to each other in a non-breakable fashion, both are signed and the signatures are stored in the blockchain.

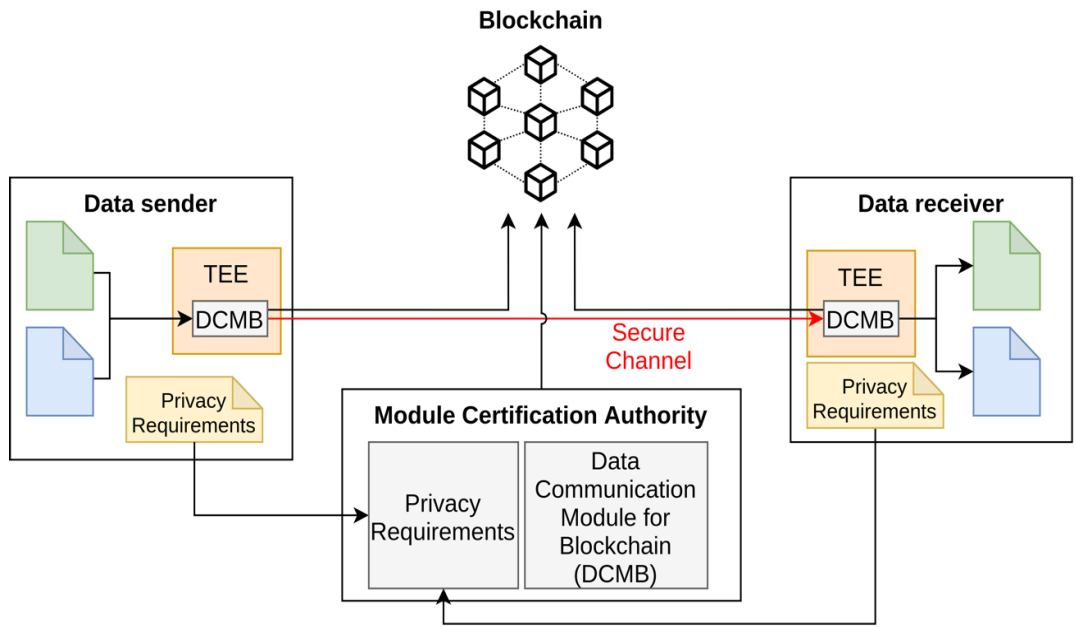

Figure 3. Blockchain Certification Authority

As seen in Fig. 3, the approach consists of three components: the DCMB data transfer via P2P communication, the Module Certification Authority for module certification and the blockchain for providing an audit trail. The Module Certification Authority consists out of a group of module validation and certification entities. It is assumed that these entities have the required domain knowledge to execute the processes of module certification well considered and carefully. To increase the availability of the certification process, while also reducing the possibility of successful manipulation, each entity executes the certification independent of the group. During the certification, each entity builds the DCMB module and validates the functionality against the confidentiality requirements and a confidentiality validation test data set. To establish consensus of the certification status, each entity participates in a voting process via smart contracts. Certification related information (e.g. validation log, hash of the source code as a reference for later audit) is available via the blockchain, to create an end-to-end auditable trail. The certification status is persisted in the blockchain and is queried during the module deployment phase. With this approach, the role of the blockchain shifts from being a distributed access management to being the canonical source of privacy module "trust".

## 4. IMPLEMENTATION OF DATA CONFIDENTIALITY USING BLOCKCHAINS IN INDUSTRY 4.0

*Data Communication Module for Blockchain (DCMB)* The DCMB provides a high degree of adaptability to be applicable for a wide variety of use cases. For example: in an IoT-centric use case, where reduction in required computation power is the top priority, the DCMB might collect data and transfers it without any further processing. If more computation power is available, the DCMB might also perform additional data validation tasks. The modular structure of the DCMB enables individual customization, depending on factors such as intended confidentiality or execution environment restrictions. To ensure compatibility between modules, may created by different entities, a standardized way of communication (e.g. OPC UA) between the modules was chosen. Unavailability of the P2P communication is detected by the modules and data is cached module internally either until the P2P communication is available, the cache is full or a predefined timeout has expired.

To prevent extraction of potentially confidential data and tamper, the modules are executed in a Trusted Execution Environment (TEE), which is available in commodity hardware as well as certain IoT devices. Before initial deployment of a DCMB module, the module needs to be certified by the module certification authority.

*Secure and Trusted Execution:* A Trusted Execution Environment (TEE) is used as protection against attacks on the DCMB modules. TEE's are available for x86 based systems through, for example, SCONE by Arnautov et al. [11] and for ARM based systems, for example, ARM TrustZone [12].

## 5. EVALUATION

To evaluate the developed concept, we define an industry 4.0-centric use case. This scenario (see Fig. 4) has been simplified to show the essential objectives of this paper: providing confidentiality protection while also providing and auditable proof of information exchange.

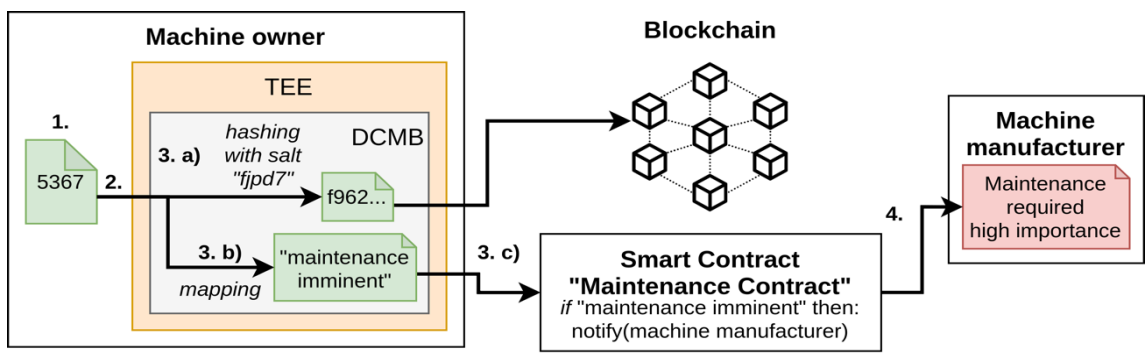

Figure 4. Industry 4.0-centric use case

The environment consists of two stockholders: a machine owner and a machine manufacturer. The machine owner wants to have an on-demand maintenance based on the number of work pieces produced. This value must be confidential to the machine manufacturer, due to financial impacts if data is released outside of the manufacturer. As a compromise between data confidentiality and maintenance service fulfillment, the value (e.g. 5367) is mapped to a value range (e.g. 5300-5400 →"maintenance imminent"). The function of the DCMB is defined by the following Algorithm 4, that describes the process of mapping the prediction value to an interval section by qualitative description of the interval.

**Algorithm 4** Machine owner
```
 1: for each 12h do
 2:     collect 5367;
 3:     val: 5367;
 4:     res: 0;
 5:     if val > 5300 and val < 5400 then
 6:         res = hash("Maintenance imminent" + "fjpd7");
 7:         send res to Blockchain;
 8:     end if
 9:     f962... = hash(5367 + "fjpd7");
10:     send: f962... and "fjpd7" to Blockchain for audit trail
11: end for
```

The smart contract "Maintenance Contract" for maintenance is defined by the following Algorithm 5, that describes the smart contract that notifies the machine manufacturer of the urgency of the pending maintenance.

**Algorithm 5** Notification Smart Contract
```
1: input: transaction
2: if transaction == "Maintenance imminent" then
3:     notify machine manufacturer with high importance
4: end if
```

The list below shows a step-by-step process of the scenario of Fig. 4 using the algorithms described above:

1. Machine owner collects sensor data
2. Machine owner transmits data to machine manufacturer via DCMB
3. DCMB steps (see Alg. 4):
   a. hashes data with salt and records hash in Blockchain adding evidence to the audit trail.
   b. maps the collected value of the machine to a value range and converts it to a qualitative value (e.g. "maintenance imminent").
   c. creates a transaction and sends it to the smart contract "Maintenance Contract" (see Alg. 5) of the blockchain.
4. Smart contract "Maintenance Contract" compares the transferred value from the machine against known conditions (hashed qualitative values) and notifies the machine manufacturer that maintenance must be performed.

To conclude, it can be said that the presented approach can be integrated into an industry 4.0 environment due to its modular design. Low resource usage, a key requirement of embedded and IoT environments, can be met by low complexity of the algorithms.

## 6. CONCLUSION

The blockchain is used to build trust between the enterprises working together. In this paper it has been shown, that an audit trail can be managed by the blockchain without having send the data through the blockchain. It even can be dynamically set up privacy channels between two enterprises preserving the data privacy. Further, it has been shown, that smart contracts can be used with hashed values protecting the value from other participants of the blockchain. This functionality has been implemented by the new designed and customizable modules: The modules are certified via a Module Certification Authority based on privacy requirements of the data collector. The process of certification is transparent for the users of the modules, since all information is stored in the blockchain.

## ACKNOWLEDGMENT

This research was funded by the European Regional Development Fund (EFRE) and by the Ministry of Science, Research and Art of the State of Baden-Württemberg, Germany.

**AUTHORS**

**Jan Stodt**, M.Sc.is a member of the Institute for Data Science, Cloud Computing and IT-security and a member of the faculty of computer science at the University of Applied Science in Furtwangen (HFU). He received his B. Sc. degree in computer science from the University of Applied Science in Furtwangen (HFU) in 2017 and his M. Sc. degree in computer science for the University of Applied Science in Furtwangen (HFU) in 2019.

**Christoph Reich** is a professor at the faculty of computer science at the University of Applied Science in Furtwangen (HFU) and teaches in the field of network technologies, programming, IT management, middleware and IT security. He has the scientific management of the HFU Information and Media Center, which consists of the departments IT, Online Systems, Learning Systems and HFU library department. As a director of the Institute for Data Science, Cloud Computing and IT-security (www.wolke.hs-furtwangen.de), his research focuses on cloud computing, QoS, virtualization and IT security.